\documentclass[twocolumn, superscriptaddress]{revtex4-2}
\usepackage[utf8]{inputenc}
\usepackage[T1]{fontenc}
\usepackage{natbib}
\usepackage{graphicx}
\usepackage{xcolor}
\usepackage{hyperref}
\usepackage{amsmath}
\usepackage{soul}

\begin{document}

\newcommand{\fig}[1]{Fig.~\ref{#1}}
\newcommand{\supinf}[1]{Appendix ~\ref{#1}}
\newcommand{\CV}{\text{CV}}
\newcommand{\new}[1]{\textcolor{red}{#1}}

\title{From asynchronous states to Griffiths phases and back:\\  structural heterogeneity and homeostasis in excitatory-inhibitory networks}
\author{Jorge Pretel}
\affiliation{%
 Departamento de Electromagnetismo y Física de la Materia and Instituto Carlos I de Física Teórica y Computacional, E-18071 Granada, Spain
}%
\author{Victor Buend\'ia}
\affiliation{%
	Deparment of Computer Science, University of T\"ubingen, T\"ubingen, Germany
}%
\affiliation{%
	Max Planck Institute for Biological Cybernetics, T\"ubingen, Germany
}%

\author{Joaquín J. Torres}
\affiliation{%
 Departamento de Electromagnetismo y Física de la Materia and Instituto Carlos I de Física Teórica y Computacional, E-18071 Granada, Spain
}%
\author{Miguel A. Muñoz}
\affiliation{%
 Departamento de Electromagnetismo y Física de la Materia and Instituto Carlos I de Física Teórica y Computacional, E-18071 Granada, Spain
}%
\date{\today}

\begin{abstract}
Balanced neural networks ---in which excitatory and inhibitory inputs compensate each other on average--- give rise to a dynamical phase
dominated by fluctuations called asynchronous state, crucial for brain functioning. However, structural disorder ---which is inherent to random networks--- can hinder such an excitation-inhibition balance. Indeed, structural and synaptic heterogeneities can generate extended regions in phase space akin to critical points, called Griffiths phases, with dynamical features very different from those of asynchronous states. Here, we study a  simple neural-network model with tunable levels of heterogeneity able to display these two types of dynamical regimes ---i.e., asynchronous states and Griffiths phases--- putting them together within a single phase diagram. Using this simple model, we are able to emphasize the crucial role played by synaptic plasticity and homeostasis to re-establish balance in intrinsically heterogeneous networks. Overall, we shed light onto how diverse dynamical regimes, each with different functional advantages, can emerge from a given network as a result of self-organizing homeostatic mechanisms.
\end{abstract}

\maketitle

\section{Introduction}

Understanding the relationship between the basic structural features of a neural network and its dynamical repertoire is crucial if one aims to construct a general modeling framework to describe brain function \cite{Deco-dynamic,Guille,unveiling}. 

Neural populations must be able to effectively process incoming signals while keeping their activity sparse and segregated from other areas involved in different tasks and, possibly, might need to transmit their outputs to other regions to integrate information at a higher level \cite{Tononi,Deco-Tononi}. This segregation-integration balance is severely constrained by  the spatial distribution and connectivity patterns of different cells \cite{Sporns-SI,Philosophical}.

Recent advances in recording techniques allow for the measurement of the activity of thousands of neighboring cortical neurons in a rather accurate manner, revealing that cells tend to fire in an irregular fashion with high variability and relatively low average pairwise correlation \cite{ecker2010decorrelated,Destexhe,Stringer-Science,Stringer-Nature}. 
This type of empirically observed collective dynamical regime, known as \textit{asynchronous state} \cite{softky1993highly,arieli1996dynamics,abeles1991corticonics,Renart} is theoretically understood as the outcome of the interplay between the opposing excitatory (E) and inhibitory (I) forces that, together, control the dynamical state of neural populations \cite{van1996, van1998,Hansel,Rubin,buendia2019jensen,Renart, alishbayli2019JN, Destexhe}.  More specifically, when the connectivity pattern is such that E/I inputs to any given neuron nearly compensate each other \emph{on average}, the activity of such a cell is dominated by fluctuations of the input rather than by its mean value, which may be below threshold. As a consequence, the resulting asynchronous state of balanced networks is characterized by sporadic firing events with high temporal variability, akin to a random Poisson process \cite{van1996, van1998,Hansel,Rubin,buendia2019jensen}.

Early theoretical/modeling approaches succeeded at reproducing this fluctuation-dominated asynchronous state considering sparse random networks and/or synaptic weights that were scaled down with network size \cite{van1996,van1998,Brunel,Hansel}. In particular, sparsity ensures that two given neurons share a negligible proportion of presynaptic neighbors and inputs, and hence their activity remains mostly uncorrelated. However, actual cortical populations are known to be densely connected \cite{fino2011dense,Fornito}. Thus, the requisite of network sparsity was later relaxed and more-refined models introduced the possibility of a \textit{dynamical} balance, by which the temporal variations in excitatory input could be rapidly tracked and compensated by variations of the inhibitory counterpart, leading to low pairwise correlations even in densely connected networks \cite{Renart}. In particular, although pairwise correlations are consistently found to be small \textit{on average} in empirical measurements of neural activity, more recent studies have reported nontrivial correlation structures in cortical circuits, such as non-monotonic dependencies with distance \cite{rosenbaum2017spatial},
heterogeneous synaptic connections \cite{Fukai,Buzsaki-log,Roxin},
and correlation patterns extending across ranges considerably larger than the typical reach of local synaptic connections \cite{dahmen2022global,Pernice}. These correlations give rise to so-called \textit{neural modes}, i.e., specific activation patterns that capture the bulk of activity variability in cortical populations, which are robustly observed, and have been proven to be crucial for brain functioning \cite{sadtler2014neural,gallego2017neural,gallego2018cortical,Stringer-Science}. Improved models account also for these features and, thus, the asynchronous-state paradigm of neural dynamics has become something of the "\emph{standard model}" in computational neuroscience \cite{rosenbaum2017spatial,Fukai,dahmen2022global,Machens,Rubin,Renart}.

In parallel to these findings, recorded neuronal activity, both in local circuits \cite{BP,hahn2010neuronal,Fontenele} and whole-brain measurements \cite{Taglia,ponce2018whole,Ponce2023}, reveals the widespread occurrence of neuronal avalanches, i.e., highly heterogeneous cascades of spontaneous activity characterized by scale-invariance in space and time \cite{Schuster,Plenz-review,Breakspear-review,RMP,Byrne,Martinello,Meshulam,Jones-Shew}. Power-law scaling in avalanche size and duration distributions constitutes a hallmark of critical behavior in stochastic models of propagating activity such as the branching process (see, e.g., \cite{Serena-branching}). Scale-free avalanching behavior appears when the system is placed at a critical point; i.e. at the edge of a transition separating a \emph{quiescent} phase, where activity decays to zero, and an \emph{active} one, where activity is self-sustained. The critical state is characterized by the divergence of physical quantities such as the correlation length and response to perturbations and has been argued to entail important functional advantages for biological systems \cite{Schuster,Sporns2004,Plenz-review,Breakspear-review,Beggs-optimize,wilting25,Byrne,RMP}. In particular, it has been shown that neural-network functional properties such as dynamic range, information transmission and information capacity can be optimized by setting the network dynamical state close to the edge of a phase transition \cite{shew2011information,shew2013functional,Sporns2004,Haimovici2013PRL,Breakspear-review,RMP,Gautam-Shew,Shew2015adaptation}.

Let us highlight that the mostly uncorrelated activity of asynchronous states stands in contrast to the huge spatio-temporal correlations near critical states \cite{dahmen2019second,wilting2019,Li-Shew}. To understand this seeming dichotomy between asynchronous and critical states, a crucial aspect of neural networks such as "network heterogeneity" needs to be further considered. Indeed, an essential feature of actual neuronal networks is that they are far from regular; thousands of interconnected cells cluster together in the cortex forming a diverse and irregular set of motifs \cite{sporns2004motifs,Sporns2004,sporns2005human,Fornito}. 
Such an inherent heterogeneity entails the emergence of local variability in neuronal densities, neuronal types, and coupling strengths giving rise to a complex structural architecture which, in turn, is at the basis of an emergent rich dynamical behavior \cite{balasubramanian2015heterogeneity,mejias-heterogeneity,Fukai,Suarez,Chen2023,Janarek2023}. The resulting complex structural and dynamical patterns constitute a substrate for the emergence of varied and intricate computational processes at the bases of cognitive functions, a richness that does not emerge in perfectly regular substrates. Structural disorder, consequently, plays a primary role in defining the emergent collective behavior and, therefore, the functional capabilities of a neural network.

From the theory of critical phenomena \cite{Binney,RMP,Sornette}, we know that the introduction of structural heterogeneity or "disorder" ---in the form, for example, of diverse coupling strengths among neurons or in their spacial localization---  may induce the emergence of \emph{rare-region effects}. Namely, the spontaneous formation of atypical local clusters with behaviour differing from the network average. Such clusters exert an influence on nearby units, leading to nontrivial collective dynamical properties of disordered systems \cite{Vojta-review,Moreira,Cafiero}. In particular, it has been reported that heterogeneous networks embedded in a physical space can display an extended region in parameter space, called a \emph{Griffiths phase}, with critical-like features such as generic power laws in avalanche distributions with varying exponents, slow dynamics, divergence in response to stimuli, etc. \cite{munoz2010griffiths,Juhasz,moretti2013griffiths,Odor-connectome,Odor2016}. 
This “stretching” of a critical point into a broad region ---besides relaxing the need for fine-tuning that characterizes criticality in ordered systems--- is likely to emerge in systems such as brain networks embedded in a physical space and with a highly heterogeneous connectivity pattern \cite{moretti2013griffiths}.
On the other hand, high degrees of heterogeneity can severely hinder the possibility of achieving a local E/I balance, so it seems that the asynchronous state itself might be compromised by the presence of heterogeneity. In particular, heterogeneity in the spatial location of excitatory and inhibitory neurons could possibly lead to a situation where recurrent activity is concentrated in excitation-dominated regions while the rest of the network might be inhibition dominated and thus remain mostly silent \cite{landau2016impact}. 

Our aim here is to investigate a very simple network model of binary neurons, such that it supports the two main paradigms of dynamical behavior in cortical networks: (i) an \emph{asynchronous phase} and an (ii) extended critical-like region, i.e. a \emph{Griffiths phase}. Employing networks with a tunable degree of
heterogeneity, we scrutinize the effects of spatial structural disorder on the emergence and possible coexistence of these two diverse dynamical regimes. In particular, we show that for a regular network embedded in an physical (Euclidean) space there is an asynchronous state with roughly equal excitatory and inhibitory inputs to each single neuron. However, this balance breaks down as the degree of heterogeneity is increased by randomly redistributing a fraction of neurons in space. Thus, spatial heterogeneity generates a state in which the activity reverberates for extended periods in local over-excited clusters while inhibition-dominated areas produce local quiescent states, i.e., a Griffiths phase. Furthermore, we study the effect of two different homeostatic mechanisms regulating the local dynamics through the modification of synaptic weights showing that both of them are able to dynamically restore the structurally broken E/I balance and generate self-organized balanced networks with a standard critical point rather than a Griffiths phase, in spite of the heterogeneous spatial distribution of neurons.

\begin{figure}[!htpb]
\includegraphics{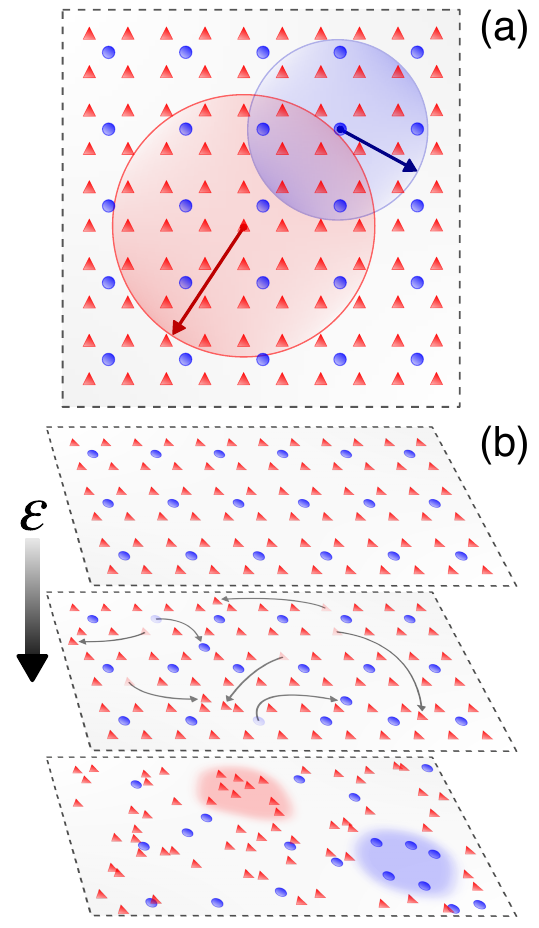}
\caption{\textbf{Network topology: 
from homogeneous to heterogeneous network architectures.} \textbf{(a)}: Spatial connectivity pattern defining a regular network (lattice) embedded in a two-dimensional substrate ($\varepsilon=0$). Note that $80\%$ of the neurons are excitatory (red) and the remaining $20\%$ are of inhibitory (blue). Each excitatory (E) or inhibitory (I) neuron projects synapses/connections to all other neurons within a given radius, $\sigma_{E/I}$ (red/blue circles for E/I neurons, respectively). \textbf{(b)}: Starting from the regular lattice, a fraction $\bf{\varepsilon}$ of all neurons is randomly relocated, thus generating regions (as the red shaded one) where excitatory neurons are over-represented as well as inhibition-dominated zones (as the blue shaded one).
As $\bf{\varepsilon}$ increases, more heterogeneous networks can be constructed (see vertical arrow). The presence of large rare regions with un-balanced connectivity patterns (marked with red and blue shaded areas in the lowest panel) influences dramatically the dynamical regime on such heterogeneous networks.}
\label{fig1}
\end{figure}

\section{Model and Methods}

We consider a simple rate model of binary neurons ---introduced in previous works \cite{larremore2014inhibition,buendia2019jensen} and similar to \cite{GS2020}--- that aims to capture the key ingredients of neural-network collective dynamics in a parsimonious way. The model consists of a network with $N$ neurons; a fraction $\alpha=0.8$ of which are excitatory (E) while the rest are inhibitory (I). 
Each neuron, $i \in [1,2,...N]$, can be either in an active, $s_{i}\left(t\right)=1$, or quiescent state, $s_{i}\left(t\right)=0$, and it projects to every other neuron within a circle of a given radius $\sigma_{k}$, with $k=E,I$ (see below and Fig.1 for a model sketch of the network architecture).

At every discrete time step $t=1,2,...$, each neuron $i$ integrates all the weighted inputs 
from its neighboring (presynaptic) neurons:
\begin{equation}
\Lambda_{i}=\frac{\gamma} {k_{i}}\sum_{j}\omega_{ij}s_{j}\left(t\right),
\end{equation}
where $\omega_{ij}$ are the corresponding synaptic weights  from neuron $j$ to neuron $i$ (positive for excitatory  $j$ neurons $\omega_{E}>0$, and negative for inhibitory ones, $\omega_{I}<0$), $k_{i}$ is the in-degree of the $i$-th neuron (used to normalize as in previous works \cite{Odor2016,Rocha2018,Barzon2022}), and $\gamma$ represents an overall coupling strength that modulates the influence of neighbouring neurons and serves as an overall control parameter.
If neuron $i$ was quiescent ($s_i=0$), it becomes active with a probability $\mathcal{P}_{i}= f(\Lambda_{i})$, 
where $f$ is a non-linear (piecewise linear) transfer function 
\begin{eqnarray}
f (\Lambda_{i})
&=& \left\{
\begin{array}{ll}
    0 & \Lambda_i<0 \\
    \Lambda_i & 0 \leq \Lambda_i \leq 1\\
    1 & \Lambda_i>1
\end{array}\right. \nonumber 
\label{eq1}
\end{eqnarray}
 Conversely, if the focal neuron was active, $s_i=1$, then it becomes quiescent with complementary probability $1-\mathcal{P}_{i}$.

The choice of $f(x)$ as a piecewise \emph{linear} function is made for the sake of simplicity; however, even if the kind of non-linearity can have an impact on some features of the system (as we discussed in Buendía \emph{et al.} \cite{buendia2019jensen}), we have verified that the forthcoming results are robust to the introduction of non-linearities in the gain function (see \supinf{app:tanh}). Similarly, without loss of generality, parameter values are fixed to $\omega_{E}=1$, $\omega_{I}=-2.5$, $\sigma_{E}=3.4$, $\sigma_{I}=2.3$ (where $\sigma=1$ is the distance between two nearby $E$ neurons in the regular network). 

The previous dynamics is first run on a regular directed network consisting of $N$ neurons embedded in an Euclidean space which, for the sake of simplicity and
without loss of generality, we consider to be two-dimensional (2D), with periodic boundary conditions. The regular grid has excitatory neurons at positions $(i,j)$, with a single inhibitory neuron per every 4 excitatory ones (as shown in Fig.\ref{fig1}). Hence, a system of size $L$ has $N=5L^2/4$ neurons. For most of the analyses, we consider $L=120, N=18000$.

This homogeneous connectivity pattern enforces that every single neuron has the same number of incoming $E$ and $I$ links (see Fig.\ref{fig1}, top). In order to study the effects of heterogeneity we also analyze the model dynamics on "perturbed" networks in which the spatial location of each neuron is allowed ---with some probability $\varepsilon$--- to shift to a new randomly selected position. Thus, $\varepsilon=0$ describes a perfectly regular network and $\varepsilon=1$ characterizes a random spatial network (Fig.\ref{fig1}), so that the relocation parameter $\varepsilon$ allows one to sample different system architectures with tunable levels of heterogeneity. 

 For the sake of completeness, the spectra of the connectivity matrices for diverse degrees of heterogeneity are shown in \supinf{app:spectra} 
\begin{figure*}[ht!]
\includegraphics{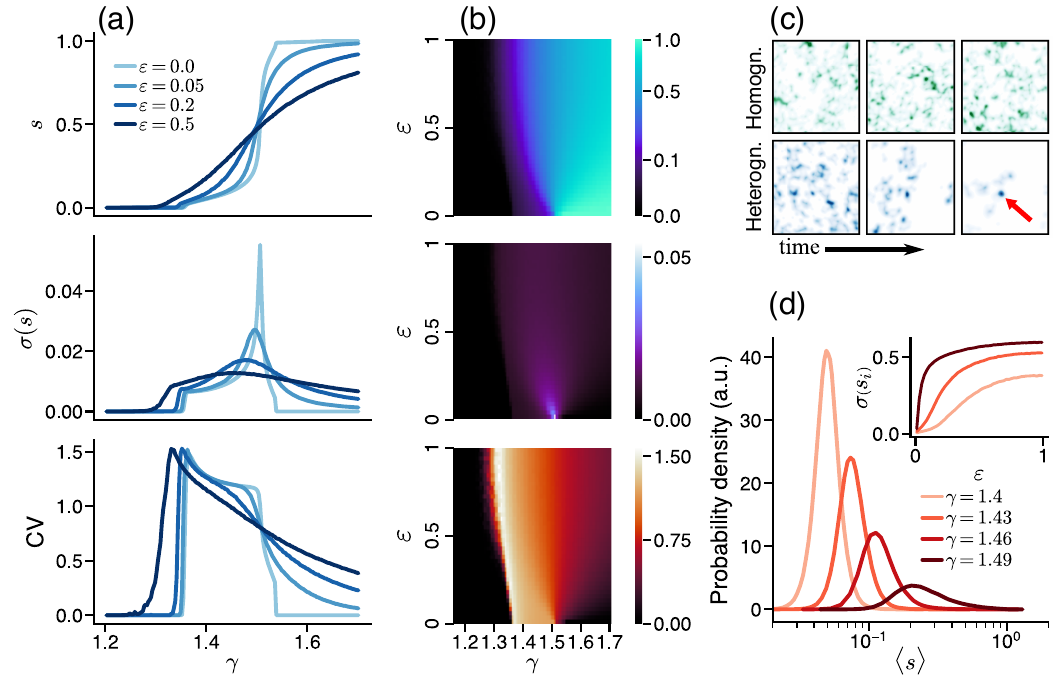}
\caption{\textbf{Phase diagram and main dynamical features.} Statistics of the activity for regular and heterogeneous networks \textbf{a)} average activity $s$ (top), activity fluctuations (standard deviation of $s$, middle), and average coefficient of variation (bottom) for different $\varepsilon$ values measured for 100 network realizations along $10^{5}$ time steps (see also \supinf{app:fss} for a finite-size scaling of the variance). \textbf{b)} average activity $s$ (top), activity fluctuations (middle) and CV (bottom) as a function of coupling $\gamma$ and disorder $\varepsilon$. \textbf{c)} slow decay of activity in the Griffiths phase (blue bottom row, $\left( \gamma,\varepsilon \right) = \left( 1.30,0.5 \right)$), displaying an active cluster that exerts influence on its neighbouring zone (red arrow), and self-sustained activity in the asynchronous phase for a regular network (green top row, $\left( \gamma,\varepsilon \right) = \left( 1.42,0.0 \right)$). \textbf{d)} Distributions of average single-cell activities can show large dispersion even for low heterogeneity levels (data for $\varepsilon = 0.05$ and increasing values of $\gamma$). \textbf{Inset}: width of the distribution of average single cell activities $\sigma \left( s_{i} \right)$ for increasing $\varepsilon$, for $\gamma=$ 1.37, 1.43 and 1.49, from bottom to top; the width of the distributions is zero at $\varepsilon=0$, and increases more sharply with $\varepsilon$ for larger synaptic weights.
}
\label{fig:main-panels}
\end{figure*}

\section{Results}

The described neuron model has been previously studied on non-spatial regular networks \cite{buendia2019jensen,Corral}. In particular, it has been shown that the presence of a set of inhibitory neurons on sparse networks gives rise to the emergence of an intermediate phase between the standard quiescent phase in which all activity eventually ceases and a phase of saturated activity in which all neurons are active, which does not exist in similar networks composed only of excitatory neurons.
This intermediate phase stems from the balance between excitation and inhibition, is dominated by fluctuations, and exhibits all the key statistical properties of asynchronous states (see \cite{buendia2019jensen,Renart}). Thus, from now on we will refer to it as the asynchronous state/phase (AS).

\subsection{Homogeneous case}
To go beyond previous results in the literature, we first verified that the same type of phase diagram arises for spatial networks when E and I neurons are regularly placed in a lattice such as the one in Fig.\ref{fig1}, i.e., for $\varepsilon = 0$. As reported in \fig{fig:main-panels}a, there is a wide range of coupling values $\gamma$ that leads to intermediate levels of self-sustained and irregular activity (see \fig{fig:main-panels}a upper panel; light-blue curve). Such an intermediate phase is separated from a quiescent one --where all activity eventually ceases--- by a critical point, $\gamma_{c1} \sim 1.365$,
as evinced by a marked increase in the variance of the overall activity (\fig{fig:main-panels}a central panel). On the other hand, the asynchronous and the saturated phases are separated by another critical point ($\gamma_{c2} \approx 1.505$), 
characterized by a more prominent peak in the activity variance (\fig{fig:main-panels}a/b central panels). Above this point, inhibition can no longer counteract excitation and the network, consequently, saturates. Timeseries illustrating the dynamics for different couplings can be found in \supinf{app:timeseries}.

The intermediate phase displays all the essential features of the asynchronous state including irregularity in the spike statistics, as quantified by the coefficient of variation (CV), defined as the quotient between the standard deviation of single-cell interspike intervals (ISI) and its mean: $\CV =\sigma_\text{ISI}/\mu_\text{ISI}$ \footnote{The ISI is measured here as the number of timesteps between deactivation and consecutive activation of a given cell.}.
Indeed, as shown in \fig{fig:main-panels}a (bottom light blue curve), the CV displays a plateau of highly variable (super-Poissonian) behavior ($\CV \approx 1.2 > 1$) which extends to the entire intermediate phase (see \fig{fig:main-panels}a/b, bottom). 

\subsection{Heterogeneous case}
When heterogeneous networks are considered ---i.e. for $\varepsilon > 0$--- the situation becomes qualitatively different as we describe in what follows. As previously summarized, the density of $E$ and $I$ neurons changes locally in space and this leads to the emergence of a heterogeneous architecture where some local regions become more prone to generate activity due to a locally increased $E/I$ ratio, while in others, activity rapidly decays given the local over-abundance of inhibitory neurons (see \fig{fig1}, bottom panel). 
This is illustrated in \fig{fig:main-panels}c; where we plot snapshots of the instantaneous state of activity in the network both for homogeneous (upper panels) and heterogeneous (bottom panels) networks, for three different times (starting with an initially saturated state). In the first case activity diffuses across the network in a rather variable way without remaining confined to any spacial location, while, in the heterogeneous case, it may remain strongly localized in some specific regions (see red arrow), around which it may reverberate for extended durations.

Heterogeneity has also an effect on the single-neuron averaged-activity distributions: these become progressively wider and start developing heavy tails as $\varepsilon$ is increased (see the inset of \fig{fig:main-panels}d where the variance of such a distribution is plotted as a function of $\varepsilon$ for different values of $\gamma$). 

On the other hand, the steady-state network-averaged activity shifts in a much less abrupt way from the quiescent to saturated phase (\fig{fig:main-panels}a, upper panel) as $\varepsilon$ grows. This is due to the 
 gradual recruiting of more and more activity-prone regions as $\gamma$ is increased (see also \fig{fig:main-panels}d).
Similarly, as heterogeneity increases the peaks in the variance are smoothed out (\fig{fig:main-panels}a, middle panel), and the CV continues to be larger than $1$ in a broad region of $\gamma$ values (see \fig{fig:main-panels}a, lower panel), being increased also for large $\gamma$-values. Thus, summing up, heterogeneity has the effect of widening the intermediate phase of moderate and fluctuating activity.

The three panels of \fig{fig:main-panels}b generalize the previous results for the activity, its variance, and the CV, respectively, by plotting them as a function of both $\gamma$ and $\epsilon$. Overall, these plots demonstrate that heterogeneity expands the intermediate phase, resulting in a more gradual transition between quiescence and saturation.

\begin{figure}[t]
\includegraphics{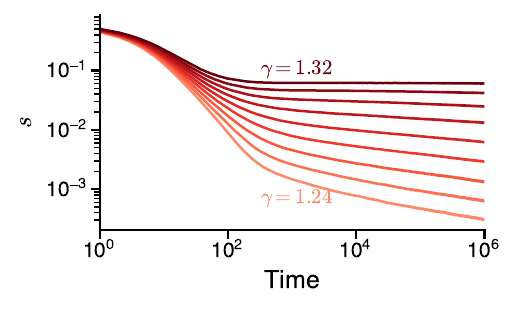}
\caption{\textbf{Griffiths phases in heterogeneous networks (I).} Generic power-law time decay of the network-averaged activity for diverse values of $\gamma=1.24, 1.25,..., 1.32$ (from bottom to top) and $\varepsilon=1.0$. The curves display a continuously varying exponent $s\sim t^{-\alpha(\gamma)}$ for an extended region in $\gamma-$space, thus defining a Griffiths phase.}
\label{fig:griffiths}
\end{figure}

We now set to look for the possible existence of a Griffiths phase for the heterogeneous-network case ($\varepsilon >0)$. As illustrated in \fig{fig:griffiths}, the overall activity (starting from a fully active initial condition) decays a
a power-law of time all across an extended region in $\gamma$-space, with continuously changing exponents, which  is one of the crucial fingerprints of Griffiths phases
\cite{munoz2010griffiths,Juhasz,moretti2013griffiths}.
As usual, the combination of (exponentially) rare large clusters in which activity can reverberate for (exponentially) large times, gives rise to this type of scale-free power-law behavior \cite{Vojta-review,moretti2013griffiths,munoz2010griffiths,Juhasz}.

\begin{figure}[h]
\includegraphics{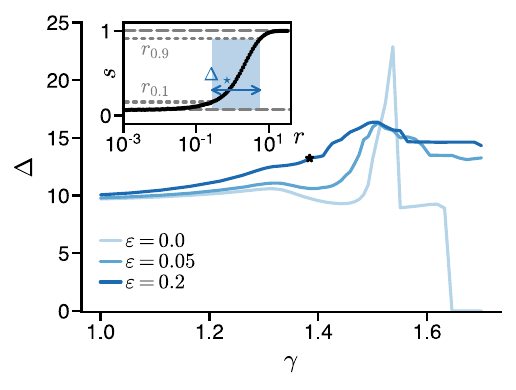}
\caption{\textbf{Griffiths phases in heterogeneous networks (II).} Dynamic range $\Delta$ (see main text) as a function of the coupling strength for different heterogeneity levels. Inset: mean activity as a function of the external input $r$ for the highlighted point in the main figure; the inset sketches how to compute the dynamic range.}
\label{fig4}
\end{figure}

Finally, we have verified that heterogeneity yields also changes in the network's response to external inputs. For this, we measured the \emph{dynamic range} $\Delta$, defined as 
\cite{kinouchi2006optimal}:
\begin{equation}
\Delta = 10\log\left( \frac{r_{0.9}}{r_{0.1}} \right)
\end{equation}
where $r_{0.1}$ and $r_{0.9}$ represent the external input intensities $\left( r \right)$ where the system displays 10\% and 90\% of the maximum possible overall activity, respectively (see \fig{fig4} inset for an illustration).  The external stimulus to each neuron $i$ consists in an additional driving rate $r$ which,
in the absence of recurrent input, causes the $i$-th neuron to activate spontaneously with a probability $f(r/k_i)$.  As shown  in Fig.\ref{fig4}, higher $\varepsilon$ values lead to an overall increase in the dynamical range of the network. The peaks around $\gamma\approx1.35$ and $\gamma\approx 1.55$ in the homogeneous case, coinciding with the transition points, are hallmarks of criticality \cite{kinouchi2006optimal}. Instead, in the heterogeneous case there is a much broader peak with larger values of $\Delta$ almost all across the parameter space, a fingerprint of a Griffiths' phase.

In summary, the asynchronous phase observed in the homogeneous case becomes a Griffiths phase with the introduction of structural heterogeneity.

\subsection{Regulatory mechanisms restore balance}

A situation in which a small subset of neurons presents exceptionally high levels of activity while most of the network remains mostly dormant does not seem to be biologically realistic. As a matter of fact, neural networks in the brain employ various strategies to maintain local balance and prevent excessive local activations. Homeostatic mechanisms ---such as threshold adaptation, synaptic scaling \cite{Turrigiano,Turrigiano-self}, rapid dis-inhibition, adaptation , and changes in intrinsic excitability --- are known to control neuronal firing rates within functionally "desirable" limits \cite{Cortes2012,gainey2017multiple,Zierenberg2018}. For instance, it has been empirically confirmed that certain types of inhibitory neurons enhance their synaptic strength to neighboring (postsynaptic) neurons if these present exceedingly high firing rates \cite{hennequin2017inhibitory}. Thus, following Landau \emph{et al.} \cite{landau2016impact}, we implemented a simplified version of the \emph{synaptic scaling} homeostatic mechanism \cite{Turrigiano,Turrigiano-self,Zierenberg2018}
in our model for heterogeneous networks. 
\begin{figure}[t]
\includegraphics{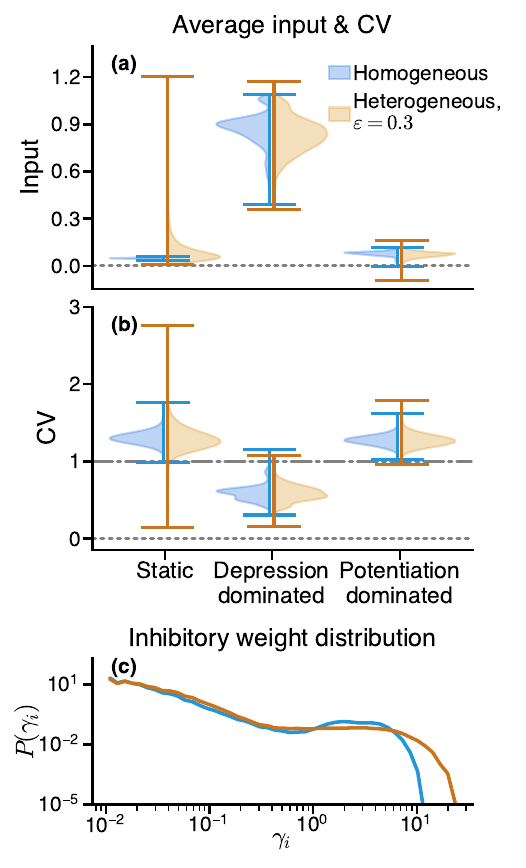}
\caption{\textbf{The effects of homeostatic plasticity.} \textbf{a} Average input to individual neurons for both homogeneous and heterogeneous networks for static (non-plastic) networks and for two choices of the ratio $\eta_p = 0.01, \eta_d = 0.1$ (depression-dominated) and $\eta_p = 0.1,  \eta_d = 0.01$ (potentiation-dominated). \textbf{b} Distribution of the coefficient of variation (CV) of the interspike-interval (ISI) of individual neurons for the same cases. Measurements were taken over $10^5$ timesteps, averaged over $10$ realizations, after $10^5$ steps. \textbf{c} Distribution of weights after $10^6$ steps in the potentiation-dominated case. Observe that the distribution is fat-tailed. All simulations were performed with $\gamma=1.4$.}
\label{fig5}
\end{figure}

More specifically, the plasticity mechanism is implemented as follows. Each active inhibitory neuron is assumed to increase the strength of its post-synaptic connections by a certain amount $\eta_p$ (\emph{potentiation}), while the synapses of inactive inhibitory neurons are decreased by an amount $\eta_d$ (\emph{depression}). In more mathematical terms, the synaptic scaling rule for the synaptic strength of inhibitory neuron $i$ is: 
\begin{equation}
    |\gamma_i(t+1)| = |\gamma_i(t)| + s_i(t)\eta_p - (1-s_i(t))\eta_d. 
    \label{adapt}
\end{equation}
Note that this scaling rule uses only local information and that to implement it, the overall synaptic weight $\gamma$ in Eq.(\ref{eq1}) has been promoted in Eq.(\ref{adapt}) to a neuron-dependent and time-dependent synaptic strength, $\gamma_i(t)$, for inhibitory neurons. On the other hand, excitatory weights are kept fixed across the network and time. Let us remark that the two plasticity parameters, $\eta_p$ and $\eta_d$, can be interpreted as the speeds of the underlying potentiation/depression process at a finer timescale.

The underlying idea behind Eq.(\ref{adapt}) is that each inhibitory neuron that is activated is likely to receive inputs from an overly excitatory region; in this way, by increasing its (inhibitory) weights it controls the excess of neighboring excitatory activity. Reciprocally, if an inhibitory neuron is inactive, weakening its weight promotes activity spreading and enhances local excitability.

To gauge excitatory-inhibitory balance in model simulations, we measured the mean input arriving at each single neuron and computed the corresponding probability distributions in the steady state (see \fig{fig5}a). One can see that, in the absence of synaptic scaling (i.e, for the "static" case, shown in the left plot), heterogeneous networks (brown curve) exhibit an input distribution which is peaked at small positive values and has a large variance and a long tail, which reveals the existence of rare strong excitatory inputs. This is to be compared with the homogeneous case (blue curves) where the distribution of inputs is almost a delta function.

On the other hand, once synaptic scaling is switched on for heterogeneous networks, they self-organize to a steady-state input distribution which depends on the parameters $\eta_p$ and $\eta_d$ of the synaptic-plasticity rule \cite{buendia2020PRR} (see \fig{fig5}a), while it is rather insensitive to variations in $\varepsilon$. In particular, we distinguish two different regimes: 
\begin{itemize}
    \item a \emph{depression-dominated} regime  $\eta_p / \eta_d < 1$  and
    \item a \emph{potentiation-dominated} regime for $\eta_p / \eta_d > 1$. 
\end{itemize}
In the depression-dominated regime, all inactive inhibitory neurons lose their weight quickly, thus leading to an excitatory-driven network in which there is a large average input with variability around it (central plots in \fig{fig5}a).
 On the contrary, if potentiation dominates, the weight of inhibitory neurons increases significantly in the presence of activity, thus compensating quickly for excitatory inputs and driving the network towards a balanced state with close-to-zero inputs and much less dispersion than in the previous case
 (see rightmost plots in \fig{fig5}a). Therefore, the type of plasticity dominated by potentiation more closely resembles what is observed in actual biological networks.

Therefore, in the presence of a potentiation-dominated homeostatic mechanism the network self-organizes to a balanced asynchronous state, whereas the inputs almost vanish on average, giving rise to a fluctuation-dominated state with irregular activity (as evinced by, e.g., the large values of the CV in \fig{fig5}b), but without large dispersion in input values.
Importantly, the degree of balance depends on the details of the adaptation mechanism; for instance, the CV is relatively small in the  depression-dominated regime. Moreover, we observed that a progressively tighter balance can be achieved by further enhancing the ratio $\eta_p/\eta_d$ with both parameters going to zero, a limit which resembles that of self-organized criticality \cite{feedback}.

Remarkably, Fig.\ref{fig5}c reports the steady-state distributions of inhibitory weights in the potentiation-dominated regime (both for homogeneous and heterogeneous networks). Such distributions exhibit a fat-tail which spans for at least three decades and are larger for the heterogeneous case, in good agreement with previous theoretical and empirical observations that systematically report broad distributions of synaptic strengths \cite{landau2016impact,Fukai,Buzsaki-log}.

\section{Discussion}

In this work, we investigated the role of structural heterogeneity on the dynamics of spatially explicit excitation-inhibition networks. For this, we have analyzed a parsimonious model of binary neurons
\cite{larremore2014inhibition,buendia2019jensen}. Previous studies of such a model had shown that the introduction of an inhibitory population on sparse random networks induces the emergence of an intermediate phase ---in between the quiescent and saturated standard ones--- of intermediate self-sustained activity, that does not exist in networks composed merely of excitatory neurons
\cite{buendia2019jensen,Corral}. Such a phase exhibits all the properties of the "asynchronous state" \cite{softky1993highly,arieli1996dynamics,abeles1991corticonics,Renart,Fukai}. 
However, previous studies did not carefully analyze spatial structure nor the possibility of structural heterogeneity and rare-region effects such as Griffith's phases (see, however, \cite{dahmen2022global}).

When space is explicitly considered, as done here, there is neuronal clustering ---i.e., nearby neurons tend to share a substantial proportion of neighbours--- which is an essential feature of actual brain networks \cite{fino2011dense}. In particular, starting from a perfectly ordered and balanced network
(the lattice shown in \fig{fig1}),
we built heterogeneous spatially embedded networks by randomly relocating neurons in the embedding space in a progressive way.
The breaking of homogeneity leads to the stochastic emergence of regions with an excess of excitatory neurons, where activity may reverberate for a long time, and regions where inhibition prevails, hindering activity propagation. As we have shown, this variability in local excitability produces smearing of the well-defined boundary between asynchronous and saturated phases, characteristic of homogeneous networks. Moreover, the resulting clustered structures lead to the emergence of a region in parameter space characterized by a generic scale-free slow relaxation, i.e. a fingerprint  of a Griffiths phase. Such a variability of dynamical timescales could be crucial for enabling and shaping temporal input integration and segregation \cite{wolff2022intrinsic}. It is noteworthy that, in the absence of external input, activity may relax to the quiescent state before the synaptic-scaling mechanism has time enough to balance the network. In this sense, the slow dynamics of the Griffith's phase might have a role in allowing synaptic scaling (or other adaptation mechanisms) to fully develop (for a discussion on the network's response in finite time, see \cite{Sahel}). 

In addition, we have also shown that structural heterogeneity promotes a more gradual recruitment of active neurons when an external input is applied, which significantly increases the overall dynamic range of the network, a highly desirable property in systems often exposed to sensory inputs that can range over several orders of magnitude. Such a network could also present advantages to store information locally and thus be efficient for working memory \cite{working-memory-mejias}.  In general, the described dynamical regimes could potentially offer functional advantages for heterogeneous networks compared to homogeneous ones. However, our primary objective here is not to examine the potential functional benefits of each dynamical regime but rather to characterize them.

In order to reconcile heterogeneity with E/I balance, we implemented ---inspired by recent work by Landau \emph{et al.} \cite{landau2016impact}--- a synaptic-plasticity rule known as "synaptic scaling" \cite{Turrigiano,Turrigiano-self}, which 
requires only local information and that allows the network to adapt its behavior to diverse dynamical regimes. In particular, we have shown that synaptic scaling of inhibitory neurons suffices to dynamically restore the network balance in heterogeneous networks. Landau \emph{et al.} found similar results for non-spatial networks of integrate-and-fire neurons \cite{landau2016impact}. Thus, our work extends the previous one by showing how synaptic scaling is able to restore the balance in spatially-embedded networks. For these, heterogeneity induces stronger effects, including the emergence of rare regions and the concomitant Griffith's phases.

Finally, depending on the relative weight of synaptic potentiation and synaptic depression, networks can attain states closer to or farther from a typical asynchronous state in a self-organized manner. This makes it possible for such adaptive networks to achieve a variety of potential dynamical regimes ---ranging from heterogeneous structures with unbalanced E/I inputs to balanced asynchronous states--- with different functional advantages for information processing. In future work, we plan to study this crossover between diverse dynamical regimes in biologically more realistic models ---including for example the possibility of oscillations, i.e. a synchronization transition \cite{LG,Liang2020}--- and to scrutinize the potential functional advantages of adaptive networks exploiting this spectrum of possibilities, ranging from heterogeneity-dominated Griffith-like states to asynchronous states. Finally, Lastly, another long-term objective of ours is to contribute to the design of neuronal networks "in vitro" by achieving a suitable balance of excitation and inhibition \cite{Yamamoto,Levina-EI}.

\section{Acknowledgements}
JP, VB, JJT, and MAM acknowledge the Spanish Ministry and Agencia Estatal de investigación (AEI) through Project of I+D+i (PID2020-113681GB-I00), financed by MICIN/AEI/10.13039/501100011033 and FEDER “A way to make Europe”, and the Consejería de Conocimiento, Investigación Universidad, Junta de Andalucía and European Regional Development Fund (P20-00173) for financial support. VB was supported by funding from Sofja Kovalevskaja Award from the Alexander von Humboldt Foundation, endowed by the German Federal Ministry of Education and Research. We would like to thank A. Levina, R. Corral, G. B. Morales, R. Calvo, C. Martorell, S. di Santo, and Gustavo Menesse for insightful discussions and comments.

\appendix
\renewcommand\thefigure{\thesection.\arabic{figure}}    
\setcounter{figure}{0}

\section{Results for another response function} \label{app:tanh}

\begin{figure}[h!]
\includegraphics{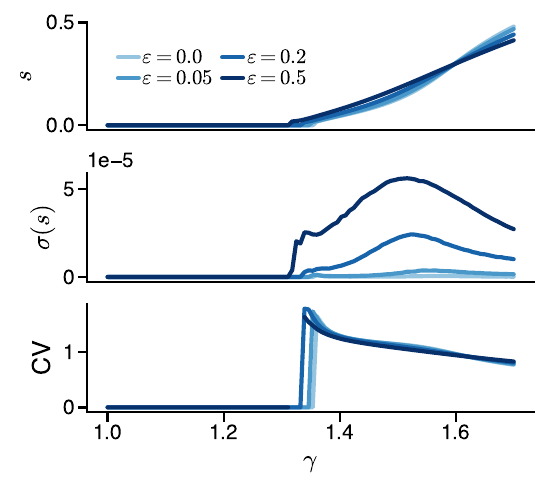}
\caption{\textbf{Results for a different response function.} Average activity $s$, variance $\sigma$ and coefficient of variation (CV) as a function of $\gamma$ for $f(x>0)=\tanh(x)$. Note that the second transition becomes continuous, so that the system does not saturate.}
\label{figapp:tanh}
\end{figure}

We have confirmed that our main results remain unchanged when varying the considered response function, even if the nature of the second transition (irregular activity to saturation) can be affected by such a choice (we refer to the Supp. Inf. of \cite{buendia2019jensen} for an in-depth discussion). In particular, \fig{figapp:tanh} shows results analogous to that of Fig.2 in the main text, but using $f(x > 0) = \tanh(x)$ and $f(x\leq0)=0$. Observe that there is no significant change. Other properties, not reported here, also remain robust against changes in the response function. 

\section{Time series} \label{app:timeseries}
\begin{figure}[h!]
\includegraphics{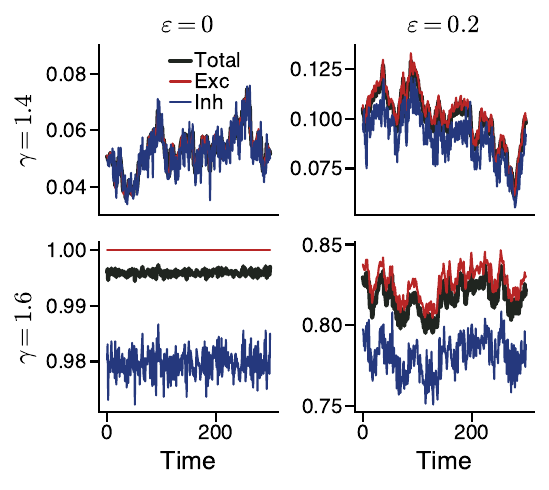}
\caption{\textbf{Time series.} Fraction of active neurons of the excitatory (red), inhibitory (blue) populations, and overall activity (black) as a function of time. Rows represent the time series for different coupling strengths, while columns are for different values of the heterogeneity parameter. }
\label{figapp:timeseries}
\end{figure}

\fig{figapp:timeseries} shows representative timeseries for two different couplings and two values of the heterogeinity parameter. Observe that in the homogeneous case (first column) $\gamma=1.4$ lies within the asynchronous irregular regime, while $\gamma=1.6$ is almost saturated. In the heterogeneous case (second column) there is no saturation and the irregular regime survives for larger values of $\gamma$.

\section{Network spectra} \label{app:spectra}

\begin{figure}[!h]
\includegraphics{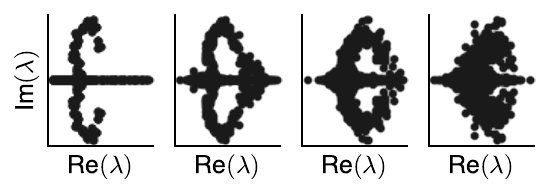}
\caption{\textbf{Matrix spectra.} Real and imaginary parts of the spectra of a matrix generated using a system size with $L=40$ (i.e., the matrix is $2000\times2000$) for increasing values of the heterogeneity parameter.}
\label{figapp:spectra}
\end{figure}

For the sake of illustration, we report the spectra of the our adjacency matrices for different values of the heterogenity parameter in \fig{figapp:spectra}. The behaviour of the system is always dominated by the largest eigenvalue, but, notice that for intermediate values of the heterogeneity there's a larger density of eigenvalues close to the leading one, leading to the slow timescales characteristic of the Griffiths' phase \cite{moretti2013griffiths}.

\section{Finite-size scaling} \label{app:fss}

\begin{figure}[!htpb]
\includegraphics{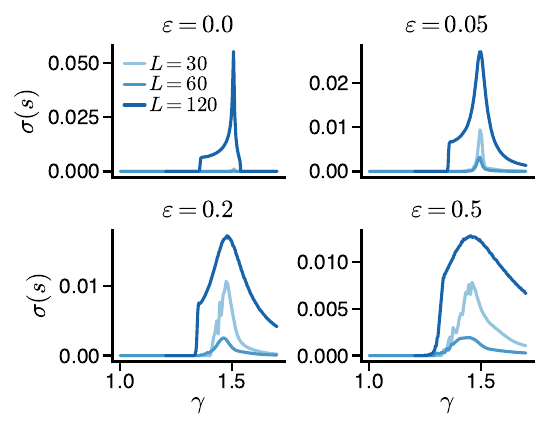}
\caption{\textbf{Finite-size scaling.} Variance of the activity as a function of the coupling strength $\gamma$ for several system sizes. Note that $N=5L^2/4$.}
\label{figapp:fss}
\end{figure}

We have performed a finite-size scaling of the phase diagram and controlled that variance of the activity tends to increase with the system size, as expected in critical phase transitions. Results are illustrated in \fig{figapp:fss}.

\vspace{10cm}
%
\end{document}